\documentclass[twocolumn,showpacs,prl,10pt]{revtex4}
\usepackage[english]{babel}
\usepackage{amsmath,amssymb}
\usepackage{times}
\usepackage{epsfig}

\begin{document}

\title {Magnetic fluctuations and resonant peak in cuprates:
a microscopic theory}
\author{I. Sega$^1$, P. Prelov\v sek$^{1,2}$ and J. Bon\v ca$^{1,2}$}
\affiliation{$^1$J.\ Stefan Institute, SI-1000 Ljubljana,
Slovenia}
\affiliation{$^2$Faculty of Mathematics and Physics, University
of Ljubljana, SI-1000 Ljubljana, Slovenia} 
\date{\today}

\begin{abstract}
The theory for the dynamical spin susceptibility within the $t$-$J$
model is developed, as relevant for the resonant magnetic peak and
normal-state magnetic response in superconducting (SC) cuprates. The
analysis is based on the equations of motion for spins and the
memory-function presentation of magnetic response where the main
damping of the low-energy spin collective mode comes from the decay
into fermionic degrees of freedom. It is shown that the damping
function at low doping is closely related to the c-axis optical conductivity. The
analysis reproduces doping-dependent features of the resonant magnetic
scattering.

\end{abstract}

\maketitle

Since its discovery in inelastic neutron scattering experiments in
superconducting (SC) YBa$_2$Cu$_3$O$_7$ \cite{ross}, the magnetic
resonance peak has been the subject of numerous experimental
investigations as well as theoretical analyses and
interpretations. The magnetic peak has been systematically followed in
YBa$_2$Cu$_3$O$_{6+x}$ (YBCO) into the underdoped regime
\cite{bour,dai,fong}, where the resonant frequency $\omega_r$
decreases while the peak intensity is increasing. Its pronounced
appearance is still related to the onset of SC, although it could
start appearing even at $T>T_c$. More recent results confirm similar
behavior in Bi2212 and Tl2201 cuprates \cite{fong1}.

Several theoretical hypotheses have been considered for the origin of
the resonant peak: that it is a bound state in the electron-hole
excitation spectrum \cite{lava}, a consequence of a novel
symmetry between antiferromagnetism (AFM) and SC \cite{deml} and that
it represents collective spin-wave-like mode induced by strong AFM
correlations \cite{morr,aban}.  There is also an ongoing debate
whether the resonant peak is intimately related to the mechanism of SC
and whether it can account for anomalies in single-electron
properties, as tested in angle resolved photoemission spectroscopy.

The scenario of a resonant mode as a collective magnetic mode seems to
correspond well to experimental facts, in particular the qualitative
development of the resonant mode with doping and its onset for
$T<T_c$. Still the status of the theory of the resonant mode, and
moreover of the magnetic response in cuprates in general, is not
satisfactory, both from the point of understanding and of the appropriate
analytical method. Relevant microscopic models, such as the Hubbard model
and the $t$-$J$ model have been so far studied in the weak coupling or
random-phase approximation \cite{lava}, neglecting strong correlations. The
latter have been considered using a Hubbard-operator
technique \cite{onuf}, and more recently within the self-consistent           
slave-boson approach \cite{brin}, self-consistent spin-fluctuation method
\cite{dahm},  as well as within the phenomenological
spin-fermion model \cite{morr,aban}.

Our aim is to develop a theory of the dynamical spin susceptibility
$\chi_{\bf q}(\omega)$ within the $t$-$J$ model. The natural
approach to analyse collective modes is the memory-function formalism
\cite{mori}.  In analogy to the previous study of spectral functions
\cite{pr} we employ the method of equations of motion (EQM) to
generate the spin dynamics and in particular to establish the
effective decay of localized spins into fermionic degrees of freedom,
which is the essential ingredient to describe the damping of the
collective mode as well as the destruction of the long-range AFM order
at finite doping.  

We study the $t$-$J$ model
\begin{equation}
H=-\sum_{i,j,s}t_{ij} \tilde{c}^\dagger_{js}\tilde{c}_{is}
+J\sum_{\langle ij\rangle}({\bf S}_i\cdot {\bf S}_j-\frac{1}{4}
n_in_j) \,, \label{eq1}
\end{equation}
on a square lattice.  In order to be close to the physics in cuprates
we take into account besides the nearest neighbor (n.n.) hopping
$t_{ij}=t$ also the  next n.n. $t_{ij}=t'$ hopping. Strong correlations among
electrons are  incorporated via
projected operators, e.g.  $\tilde{c}^\dagger_{is}= (1-n_{i,-s})
c^\dagger_{is}$, which do not allow for double occupancy of sites, and
in $J \ll t $ where for cuprates we assume furtheron $J = 0.3t$.

Within the memory function approach of Mori \cite{mori} to dynamical response
functions, which we here pursue, the dynamical spin susceptibility $\chi_{\bf
q}(\omega)$ can be expressed as
\begin{equation}
\chi_{\bf q}(\omega)=\frac{-\eta_{\bf q}}{\omega^2+\omega 
M_{\bf q}(\omega) - \omega^2_{\bf q}}\,, \label{chiq}
\end{equation}
in which form it is particularly well suitable for the analysis of
coherent collective magnetic response, as manifest in the resonant
peak in cuprates. Here, $\omega_{\bf q}$ is related to the dispersion
of the collective mode provided that the mode damping is small enough,
i.e. $\gamma_{\bf q}\sim M^{\prime\prime}_{\bf q}
(\omega_{\bf q}) <\omega_{\bf q}$. In the opposite case, we are
dealing with an overdamped mode, as seems to be generally the case for
the magnetic response near the AFM wavevector ${\bf q} \sim {\bf
Q}=(\pi,\pi)$ in the normal state of cuprates.

In order to evaluate the quantities entering Eq.(\ref{chiq}) we follow
the formalism of memory functions \cite{mori}, defining the scalar
products and projections in terms of static response functions $(A|B)=
\chi^0_{AB}= -\langle
\!\langle A^\dagger;B \rangle \!\rangle_{\omega=0} $, and the action 
of the Liouville super-operator ${\cal L}A=[H,A]$. Within this
framework we can  expres
\begin{eqnarray}
\eta_{\bf q}=({\cal L}S^z_{\bf q}|{\cal L}&S^z_{\bf q})=
\langle [S^z_{-\bf q}\,{\cal L}S^z_{\bf q})]\rangle\,,
\quad \omega^2_{\bf q}=\eta_{\bf q}/\chi^0_{\bf q} \,,
 \nonumber \\ 
M_{\bf q}(\omega)=&\!\!(\widetilde
Q{\cal L}^2S^z_{\bf q}| [{\cal L}_{\widetilde Q}-
\omega ]^{-1}|\widetilde Q{\cal L}^2S^z_{\bf q})\,/\eta_{\bf q}\,, 
\label{eq4}
\end{eqnarray}
where ${\cal L}_{\widetilde Q}=\widetilde Q {\cal L}{\widetilde Q}$ is the
projected  Liouville super-operator, $\chi^0_{\bf q}=\chi_{\bf q}(\omega=0)$
is the static susceptibility and the projectors $\widetilde Q=Q'Q$ are given
by  $Q=1-|S^z_{\bf q})[\chi^0_{\bf q}]^{-1}(S^z_{\bf q}|$ and
$Q'=1-|{\cal L}S^z_{\bf q}))[\eta_{\bf q}]^{-1} ({\cal L}S^z_{\bf q}|$,
respectively.

In order to proceed we write down equations of motion (EQM) for the
spin operators $S^z_{\bf q}$. By evaluating ${\cal L}S^z_{\bf q}$ it is
straightforward to explicitly express $\eta_{\bf q}$ in Eq.(\ref{eq4}) as
\begin{eqnarray}
\eta_{\bf q} =&\frac{1}{4N}\sum_{{\bf k},s} 
[\epsilon^0_{{\bf q}+{\bf k}}+\epsilon^0_{{\bf q}-{\bf k}}
-2 \epsilon^0_{\bf k} ] 
\langle\tilde c^{\dagger}_{{\bf k}s}\tilde c_{{\bf k}s}\rangle \nonumber \\
+& \frac{1}{2N}\sum_{\bf k} [J_{{\bf q}+{\bf k}}+J_{{\bf q}-{\bf
k}}-2J_{\bf k}] \langle S^+_{\bf k}S^-_{\bf-k}\rangle\,, \label{etaq}
\end{eqnarray}
where $\epsilon^0_{\bf k}$ is the 'free' band dispersion following
from tight-binding hopping $t_{ij}$. We are interested in the vicinity
${\bf q}\sim {\bf Q}$. We note that $\eta_{\bf q}$ is closely
related to the internal energy, i.e. $\eta_{\bf Q} \sim -\langle H
\rangle$.  This indicates that $\eta_{\bf q}$ is not strongly
dependent neither on temperature $T$ nor on ${\bf q}$, while the doping
dependence $c_h=1-\langle n_i\rangle$ (at low doping) 
$\eta \sim a c_h |t| + bJ$ is as well modest,
where $a\!\sim \!b\sim \!1$, as inferred e.g. from numerical studies
of the model.

The evaluation of ${\cal L}^2 S^z_{\bf q}$ is also straightforward, but
more tedious. It is convenient to separate the action ${\cal L}^2
={\cal L}_t^2+ {\cal L}_I^2 +{\cal L}_J^2$ with ${\cal L}_I^2=[{\cal
L}_t,{\cal L}_J]_+$, involving different powers of kinetic and
exchange terms, respectively.  In the following analysis we assume
that for the damping function $\Gamma_{\bf q}(\omega)=M^{\prime\prime}_{\bf
q}(\omega)$ at ${\bf q} \sim {\bf Q}$, at low $T\sim 0$
and in the doping regime of interest, i.e. of low to intermediate
doping, the essential term is ${\cal L}_t^2$.  Namely, we presume that
at least in the normal state the damping
$\Gamma_{\bf q}(\omega)$
approaches a constant $\gamma_{\bf q}$ for $\omega \to 0$ 
as in a Fermi liquid (although
anomalous). Such a damping can arise only from the coupling to
fermionic degrees of freedom, which then has to involve $H_t$ term.

One can give more arguments in support of our assumption. In an
undoped system - AFM, only ${\cal L}_J^2$ is effective. However, it is
well known that at $T=0$ such a term gives a damping $\gamma_{\bf q}
\propto \tilde q^2$ where $\tilde {\bf q}={\bf q}-{\bf Q}$, as well as
$\Gamma_{\bf Q}(\omega) \propto \omega^2$\cite{bech} . Hence it leads to
underdamped and well defined AFM magnon excitations for $\tilde q \to
0$, the origin being in the phase space of low-lying spin excitations
restricted to $\tilde q \sim 0$. The argument about the ${\cal L}_J^2$
term can be extended to a doped system. Assuming the Fermi liquid form
$\chi_{\bf q}^{\prime\prime}(\omega) \propto \omega$ a mode-coupling
treatment here would again yield 
$\Gamma_{\bf q}(\omega) \propto \omega^2$.
The role of ${\cal L}_I^2$, on the other hand, cannot be apriori
neglected at $\omega \to 0$.  However, within the same decoupling described
below we get a vanishing contribution of diagonal ${\cal L}_I^2
S^z_{\bf q}$ to $M_{\bf q}(\omega)$.

We should stress that the evaluation of ${\cal L}_t^2 S^z_{\bf q}$
requires explicit consideration of the projections of fermionic
operators in $H_t$. In the site-representation we can express
\begin{eqnarray}
&{\cal L}^2_{t}{\bf S}_j =-\sum_{k} t^2_{jk}({\bf S}_j-{\bf S}_k){\cal
P}_{jk} + \nonumber \\
&+\sum_{kls} t_{jk} [t_{kl} {\bf S}_j {\cal T}_{jk}  
\tilde c^{\dagger}_{js} \tilde c_{ls} - t_{jl} {\cal T}_{jl}
\tilde c^{\dagger}_{ls} \tilde c_{ks} {\bf S}_k]+{\mathrm H.c.} \label{siteqm}
\end{eqnarray}
Here ${\cal T}_{ij}=n_i(1-n_j)+{\cal P}_{ij}$ and ${\cal P}_{ij}=n_i
n_j/2+2{\bf S}_i\!\cdot\! {\bf S}_j$ is the spin-interchange operator.
Complicated form of Eq.(\ref{siteqm}) reflects the well-known involved
nature of correlated hopping in a strongly correlated system,
i.e. with a reshuffling of spins along the hole path. Since the main
goal is to get the coupling to nonlocal fermionic degrees, in EQM we
replace operators ${\cal P}_{ij},{\cal T}_{ij}$ by
their thermal averages $P_{ij}, T_{ij}$,
respectively,  leading to an effective hopping renormalization in
Eq.(\ref{siteqm}).  Note that for the n.n. hopping, $T_1$ represents
essential reduction, since in a Ne\' el state one would get $P_1=0$
and only in heavily doped system $ P_1\sim (1-c_h)^2/2$. For the regime of
interest, i.e., $0.1<c_h<0.25$ one can on the basis of numerical results for the
$t$-$J$ model simplify $T_1 \sim c_h$.  We have to apply to
Eq.~(\ref{siteqm}) also the projector $\widetilde Q$. To the lowest
order this implies that the projected operator does not contain
explicitly the initial operator $S^z_{\bf q}$ itself. So we get
\begin{eqnarray}
&\widetilde Q{\cal L}^2_t S^z_{\bf q} \sim \frac{1}{2 \sqrt{N}}
\sum_{{\bf k} s} w_{{\bf kq}} s \tilde c^{\dagger}_{{\bf k} s} \tilde
c_{{\bf k}+{\bf q},s}\, , \label{eqm} \\
&w_{{\bf kq}} = (\epsilon^0_{\bf k}-\epsilon^0_{{\bf k}+{\bf q}})
(\tilde\epsilon^0_{\bf k}-\tilde\epsilon^0_{{\bf k}+{\bf q}})-
\zeta_{\bf q}\, , \nonumber
\end{eqnarray}
where $\tilde\epsilon^0_{\bf k}$ is defined with renormalized hopping
parameters ${\tilde t}_{ij}\approx t_{ij}T_{ij}$ whereas $\zeta_{\bf
q}$ is determined by the condition $\sum_{\bf k}w_{\bf kq}=0$.  We
note that within an analogous approximation ${\cal L}^2_I S^z_{\bf q}=0$.

Eq.(\ref{eqm}) represents a decay of spin variables into fermions in a
doped system, so performing a decoupling (in a normal state) we get
for the damping in the lowest approximation
\begin{equation}
\Gamma_{\bf q}(\omega) =\frac{1}{2\eta_{\bf q}}  \int
\frac{d\omega^\prime}{\omega} [f(\omega-\omega^\prime)-f(\omega^\prime)] 
R_{\bf q}(\omega,\omega^\prime)\, ,
\label{m2p}
\end{equation}
with 
\begin{equation}
R_{\bf q}(\omega,\omega^\prime)=\frac{\pi}{N}\sum_{\bf k} w^2_{\bf kq}
A_{\bf k}(\omega^\prime) A_{{\bf k}+{\bf q}}(\omega-\omega^\prime) \, ,
\label{rcoh} 
\end{equation}
where $A_{\bf k}(\omega)$ are electron spectral functions.
At low doping an alternative decoupling of fermionic operators directly in the site
representation, Eq.(\ref{siteqm}), neglecting the coherence between
different sites might be more appropriate,
considering the fact that in underdoped systems spectral functions exhibit
pronounced  incoherent behavior. This would yield
\begin{equation}
\widetilde R_{\bf q}(\omega,\omega^\prime)\approx \pi\zeta_{\bf q}^2
{\cal N}(\omega^\prime) {\cal N}(\omega-\omega^\prime) \, , \label{ricoh}
\end{equation}
where ${\cal N}(\omega)=(2/N)\sum_{\bf k}A_{\bf k}(\omega)$ is the
electron density of states (DOS). The form (\ref{rcoh}, \ref{ricoh})is particularly appealing since by
Eq.(\ref{m2p}) the damping becomes proportional to the c-axis
conductivity, i.e.  $\Gamma_{\bf q}(\omega)
\propto\sigma_c(\omega)$, which can be well represented within the
same form \cite{prs}.

Before proceeding to discussion of $\Gamma_{\bf q}(\omega)$ we note that
the theory requires an additional input, i.e. in Eq.(\ref{chiq}) we
need either $\omega_{\bf q}$ or $\chi^0_{\bf q}$.  A possibility is
to fix unknowns with the sum rule in the paramagnetic phase
\begin{equation}
\frac{1}{\pi}\int_0^\infty d\omega ~{\rm cth}\frac{\omega}{2T}
\chi^{\prime\prime}_{\bf
q}(\omega)= \langle S^z_{-{\bf q}} S^z_{\bf q}\rangle = C_{\bf q}\, ,
\label{eqsum}
\end{equation}
where in addition $(1/N)\sum_{\bf q} C_{\bf q}=(1-c_h)/4$. Static
correlation functions $C_{\bf q}$ are rather well known within the
$t$-$J$ model \cite{sing}, in particular one can express $C_{\bf
Q}\propto \xi^2$ where $\xi$ is the AFM correlation length, determined
also experimentally in La$_{2-x}$Sr$_x$CO$_4$ \cite{birg}, with
$\xi \propto1/\sqrt{c_h}$.

In the present study we assume some simple forms for spectral
functions $A_{\bf k}(\omega)$ although there exists also an analogous
approach to fermion dynamics, yielding as well pseudogap (PG)
features in $A_{\bf k}(\omega)$ from the coupling to spin degrees
\cite{pr}.  The simplest assumption in the normal state is to insert
some effective coherent band crossing the Fermi energy, i.e.
$A_{\bf k}(\omega) \sim Z_{\bf k} \delta(\omega - \epsilon^{eff}_{\bf
k})$.
Such a form yields at $T\to 0$  $\Gamma_{\bf Q}(\omega \to
0) \sim \mathrm {const}$ from $R$, Eq.(\ref{rcoh}), as well as from
expression (\ref{ricoh}), provided that the Fermi surface crosses the
AFM zone boundary. The doping dependence enters $\Gamma$ in
several ways. First, there should be an overall proportionality to
$c_h$ which is evident from the relation of $\Gamma$ with
$\sigma_c$. In an
effective band picture the latter can arise from vanishing QP weight
$\bar Z \propto \sqrt{c_h}$ assuming that the effective band width
$\widetilde W \sim {\tilde a} c_h |t|+{\tilde b} J$ does not vanish
for $c_h \to 0$.  $\eta$ and $T_{ij}$ are less $c_h$-dependent in the
regime of interest. We can estimate the size of normal-state damping
as $\gamma_{\bf Q} \sim \pi\zeta^2_{\bf Q}\bar Z^2 /2\widetilde W^2\eta_{\bf Q}$,
which at low doping can be quite small $\gamma_{\bf Q} \ll t$.
Nevertheless, from available numerical data we estimate $\gamma_{\bf Q}$
still  too large for an underdamped collective mode 
at $\omega_{\bf Q}$ to exist.

In order to get un underdamped resonant mode at $\omega_r$ one needs
a depleted damping $\Gamma_{\bf Q}(\omega_r)$. The
latter can evidently arise in the SC state, $T<T_c$, from the SC gap. We
can model this by introducing into Eqs.(\ref{rcoh},\ref{ricoh}) an effective
d-wave gap $\Delta_{\bf k}=\Delta_0(\cos k_x-\cos k_y)/2$ via $A_{\bf
k}(\omega)$ and ${\cal N}(\omega)$.  Due to the broken symmetry we
include in Eq.~(\ref{rcoh}) also the anomalous spectral functions
$F_{\bf k}(\omega)$ \cite{morr}. The SC gap eventually leads to
the vanishing of $\Gamma_{\bf Q}(\omega<\omega^*_{\bf Q})=0$
where $\omega^*_{\bf Q} \sim 2 \Delta_{{\bf k}^*} <2 \Delta_0$ while
${\bf k}^*$ is the position of the 'hot spot' along the AFM zone
boundary.

Such an analysis with a SC d-wave gap is particularly appropriate
for the situation close to optimum doping. Results for 
$\Gamma_{\bf q}(\omega)$ for several ${\bf q}$ along the zone
diagonal are presented in Fig.~1a. Parameters are chosen so as to
correspond to optimum doping $c_h \sim 0.2$, i.e.  effective band with
$\tilde t=0.3t$, $\tilde t^\prime=-0.1 t$, $\bar Z=0.4$, and we
take $\Delta_0 \sim 0.1t$.  We note that the damping with a single step at ${\bf
q}={\bf Q}$ 
develops two steps for ${\bf q} \ne {\bf Q}$ and
the threshold $\omega^*_{\bf q}
\to 0$ closes for $\tilde q =|{\bf q}-{\bf Q}|>q^* \sim 0.3$
\cite{morr}.  This is a mechanism for the onset of strong damping of
the collective mode for $\tilde q>q^*$ and its disappearance as the
resonant character emerges. Another important feature is the 'normal
state' damping $\gamma_{\bf Q}=\Gamma_{\bf
Q}(\omega>\omega_{\bf Q}^*)$ calculated in this approach explicitly. We note
that $\gamma_{\bf Q}\gg \omega_r$, preventing any coherent feature in
the normal state.

\begin{figure}[htb]
\centering
\center{\epsfig{file=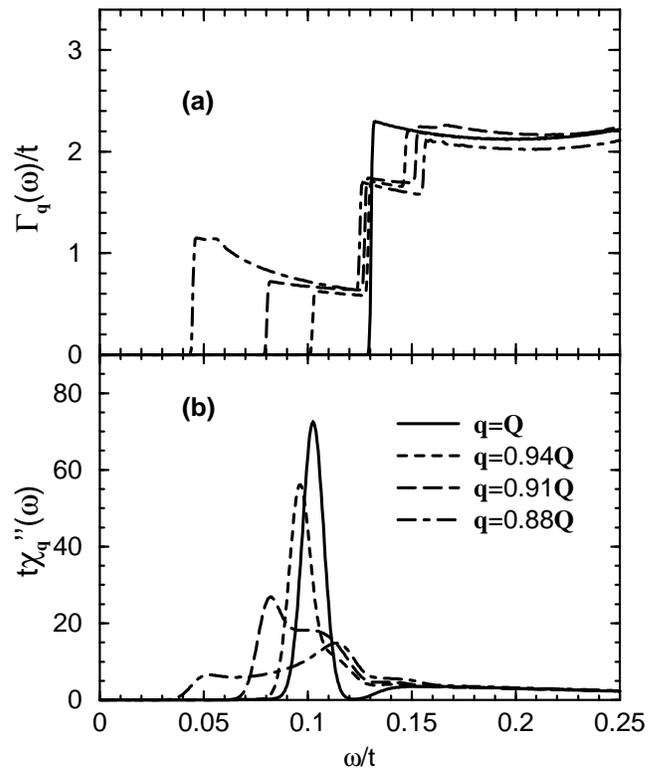,width=85mm}}\\
\caption{(a) Damping function $\Gamma_{\bf q}(\omega)$ at
optimal doping $c_h\sim 0.2$ for various $\bf q\parallel{\bf Q}$ for a d-wave SC, 
(b) corresponding spin response $\chi _{\bf q}^{\prime\prime}(\omega)$. Note that 
$t\sim 400$meV. The resonance peaks are artificially broadened with
$\delta=0.01t$.}
\label{fig1}
\end{figure}

The corresponding $\chi^{\prime\prime}_{\bf q}(\omega)$ is presented
in Fig.~1b.  We choose $\omega_{\bf q}\sim 0.38t$ such as to yield $C_{\bf q}\sim0.4$,
being consistent with results within the $t$-$J$ model for
intermediate doping \cite{sing}. At ${\bf q}={\bf Q}$ the resonant
mode is undamped since $2\Delta_0 > \omega_r$.  It should be stressed,
however, that due to quite large 
$\Gamma_{\bf Q}(\omega>\omega^*_{\bf Q})$ the resonant frequency is significantly
renormalized
\begin{equation}
\omega_r=\omega_{\bf Q} \bigl[1+M^\prime_{\bf Q}(\omega)/\omega|_
{\omega=\omega_r} \bigr]^{-1/2},
\label{peakp}
\end{equation}
while its intensity is also reduced $I_{\bf Q} \sim C_{\bf
Q}\omega_r/\omega_{\bf Q}$.  The rest of the spectral weight is
distributed over a shallow but very broad continuum.  On
the other hand, moving away from ${\bf Q}$ the mode  gets
overdamped and merges with a broad continuum for $\tilde q>q^*$. In
Fig.~1b we as well observe a downward dispersion of the resonant
peak consistent with experiments \cite{bour1,chub}.

Analysing the regime of low doping it appears more appropriate to use
the incoherent approximation, Eq.(\ref{ricoh}). It is crucial that the
normal-state damping $\gamma_{\bf Q}$ also decreases with doping,
scaling approximately as $\propto c_h$. On the other hand, it is
rather clear that for underdoped cuprates the experimental data in the
SC phase cannot be explained with a single gap only. Neutron
scattering results for $\chi^{\prime\prime}_{\bf Q}(\omega)$ in the
underdoped YBCO \cite{bour} indicate on the appearance of the
resonance at $\omega_r $ at $T<T_c$, possibly even at $T>T_c$
\cite{dai}. However, in contrast to optimum doping the resonant mode
at $\omega_r$ is quite damped. At the same time, the pronounced
shoulder below the resonance, i.e. $\omega_c<\omega_r$, appears
\cite{bour}.  The drop in $\chi^{\prime\prime}_{\bf Q}(\omega<\omega_c)$
can be again interpreted with a coherent SC gap in
$\Gamma_{\bf Q}(\omega)$, but with substantially diminished
gap $2\Delta_c<\omega_r$. Since the 'normal' damping is still too
large, i.e. $\gamma_{\bf Q}>\omega_r$, we need to assume also the
appearance of a pseudogap  in DOS
below $\omega\sim\omega_p$ for
$T<T^*$ with $T^*\geq T_c$. This is well consistent with the behavior
of $\sigma_c(\omega)$ in underdoped cuprates \cite{tama} where the
pseudogap appears at $T<T^*$.
\begin{figure}[htb]
\centering
\epsfig{file=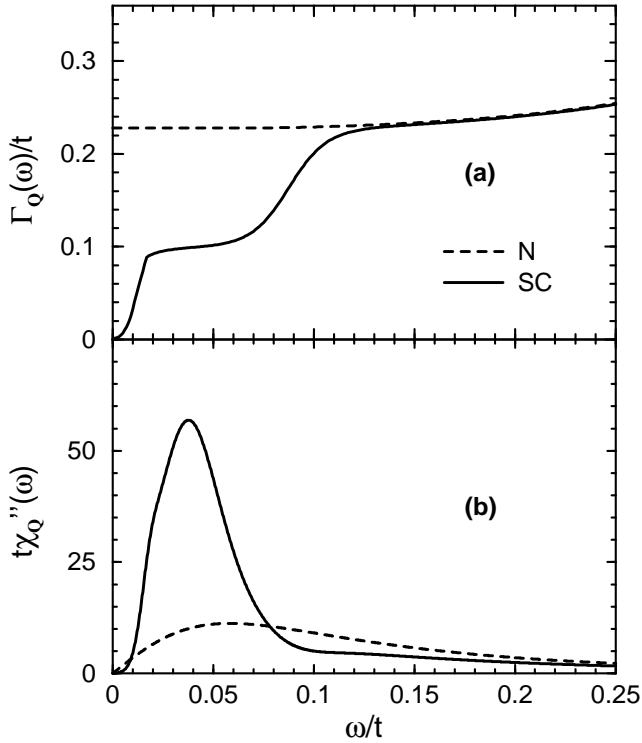,width=85mm}
\caption{(a) $\Gamma_{\bf Q}(\omega)$ for underdoped $c_h\sim 0.1$.
Dashed curve: normal state regime, $T\sim T^*\sim 0.05t$. Full curve: SC regime. Here
damping is assumed proportional to $\sigma_c(\omega)$ with an SC gap
added below $\omega_c \sim 0.01t$. (b) Corresponding
$\chi^{\prime\prime}_{\bf Q}(\omega)$.} 
\label{fig2}
\end{figure}
In Fig.~2a we present characteristic $\Gamma_{\bf Q}(\omega)$, calculated at
$T \sim 0$ for $c_h \sim 0.1$ with corresponding $\Delta\sim 0.01t$ and
calculated $\eta_{\bf Q}\sim 0.25t$ whereas $t{\cal N}(\omega>\omega_p)\sim
0.15$, as inferred from numerical results for the $t$-$J$ model \cite{jjpp}.
An additional PG reduction for $\omega<\omega_p \sim 0.1t$ is assumed,
consistent with experimental data \cite{tama} together with an estimate of
the DOS within the pseudogap $\,t{\cal N}(\omega <\omega_p)\sim 0.1$
\cite{jjpp}. The resulting $\chi^{\prime\prime}_{\bf Q}(\omega)$,
corresponding to $C_{\bf Q}\sim 1.0$, is shown in
Fig.~2b. In the normal state $\chi^{\prime\prime}_{\bf Q}(\omega)$ is
overdamped. Still, a substantial part of the sum rule is exhausted in the
window $\omega<\omega_{\bf Q}$. It should be also pointed out that in the
normal state the role of $T>0$ is essential via the sum rule (\ref{eqsum})
leading to $\omega_{\bf Q}(T)$ shifting with $T$. From Fig.~2b it is evident
that also in the SC state several features are different when compared to
larger doping in Fig.~1b: a) The (spin) gap shoulder appears for
$\omega<\omega_c$ below which there are no spin excitations.  b) The resonant
peak is damped even for $T<T_c$, but still underdamped. c) The spin response
and the sum rule for $\chi^{\prime\prime}_{\bf Q}(\omega)$ are nearly
exhausted within $\omega<\omega_p$.  Since in the underdoped regime 'normal'
$\gamma_{\bf Q}$ is also reduced, peak position given by Eq.(\ref{peakp}) is
not significantly renormalized and $\omega_r \sim \omega_{\bf Q}$.
Consequently, for the underdamped mode using the sum rule, we obtain
$\omega_r
\sim \eta/C_{\bf Q} \propto 1/\xi^2 \propto c_h$.

To summarize, our analysis of the dynamical spin response and
resonance peak in cuprates within the memory function approach can
qualitatively, and at low doping even quantitatively, reproduce the
spectra as measured in  neutron scattering experiments
\cite{ross,bour,dai,fong}. The central point of the present 
theory is the evaluation of the damping function $\Gamma_{\bf
q}(\omega)$, using the EQM and considering the decay of
spin fluctuation into electron-hole excitations as the dominant
process. In the normal state we obtain a large damping $\gamma_{\bf
Q}>\omega_{\bf Q}$ increasing with doping, leading generally to an
overdamped AFM collective mode. Still it is important to realize that the
renormalization due to $T_{ij}\ll 1$ in Eq.(\ref{eqm}) is essential.
Namely, without this reduction the damping would be much
too large and in particular it would prevent the matching of the sum
rule Eq.(\ref{eqsum}) with any pronounced short range AFM order 
$C_{\bf Q} \gg 1/4$. 

Addressing the resonance peak in the SC state, we find that in optimally
doped samples it can arise only inside the frequency gap $\sim
2\Delta_0$, being strongly renormalized in comparison to the
characteristic frequency $\omega_{\bf Q}$ and also reduced in the
intensity. On the other hand the incoherent part of spin
fluctuations extends over a broad frequency range $\sim \widetilde W$ 
and for the chosen parameters  accounts for $\sim 80 \%$ of the integrated
intensity $\int \chi^{\prime\prime}_{\bf Q}(\omega)d\omega$. 
In the underdoped  (weakly  doped) case, however, the
sum rule at $T=0$ is almost completely exhausted within the peak
width.  As a consequence we obtain in this regime $\omega_r \propto c_h$,  
quite consistent with experiments \cite{bour}. Moreover, in contrast to
optimum doping with a single SC gap, in underdoped
samples two scales seem to play the role in the SC state.
The proportionality of damping to
$\sigma_c$ would then indicate a gradual closing of the pseudogap leading to a
crossover into an overdamped situation.   
More detailed analysis will be presented elsewhere.

Authors acknowledge the support of the Ministry of Education, 
Science and Sport of Slovenia.

\end{document}